\documentstyle[aps,twocolumn,prl,epsf]{revtex} 
\begin{document}

\draft
\twocolumn[    

\hsize\textwidth\columnwidth\hsize\csname @twocolumnfalse\endcsname    

\title{Effect of Particle-Hole Asymmetry on the Mott-Hubbard Metal-Insulator Transition}

\author{D. O. Demchenko, A. V. Joura, and J. K. Freericks}
\address{Department of Physics, Georgetown University, Washington, DC 20057-0995, U.S.A.}

\date{\today}
\maketitle

\widetext
\begin{abstract}
The Mott-Hubbard metal-insulator transition is one of the most important
problems in correlated electron systems.  In the past decade, much progress
has been made on examining a particle-hole symmetric form of the transition
in the Hubbard model with dynamical mean field theory where it was found that
the electronic self energy develops a pole at the transition.
We examine the particle-hole asymmetric metal-insulator transition
in the Falicov-Kimball model, 
and find that a number of features change when the
noninteracting density of states has a finite bandwidth. 
\end{abstract}

]      

\narrowtext

The Mott-Hubbard metal-insulator transition \cite{mott,hubbard} 
is a classic example of the 
physics of strongly correlated electrons. The physical mechanism of the 
transition arises from a local Coulomb repulsion $U$ that forbids double
occupancy of the electrons, creating an insulator when there is one particle
per site (on average). 
Experimentally it is found in a variety of materials, including 
many transition metal compounds (MnO, NiO, NiS, YBa$_2$Cu$_3$O$_6$, etc.) 
for which band structure calculations
severely underestimate the band gap or yield a metallic solution \cite{tmo}.

In the Hubbard model~\cite{hubbard}, a number of approximations have been employed
that assume either the metallic state and head toward the 
insulator~\cite{hubbard,gutzwiller,ipt},
or vice versa~\cite{hubbard,brinkman_rice}.  Often it is difficult to develop an approximate theory
that is able to describe both the weakly correlated Fermi-liquid phase and
the strongly correlated insulator, hence most approximate methods yield only
limited information about the transition.  Much progress has been made with
dynamical mean field theory (DMFT)~\cite{metzner_vollhardt}, where fundamental 
questions such as,
is the transition continuous or discontinuous, does the Fermi-liquid metal
survive up to the transition or do non-Fermi-liquid metallic phases intervene, 
do metastable phases exist, and so on have been
analyzed in great detail.  However,
the numerics are quite complicated and delicate, 
because the different phases are separated by very small energies.  Hence
there has been much controversy about the answers to these questions and about
the details of the Mott-Hubbard transition~\cite{QMC_Rozenberg_Kotliar,%
EDA_Caffarel,PSCA_Moeller,AntiKotliar,Noack_Gebhard,AntiKotliar2,%
NRG_Bulla,Byczuk1}.

Some of these problems arise from the limit of infinite dimensions, where
the noninteracting density of states (DOS) on a hypercubic lattice is
a Gaussian, which has an infinite bandwidth.  This means that the DOS can
only vanish (and thereby yield an insulator at $T=0$) when the self energy 
diverges.  Since this occurs only at the single point of an isolated pole, the 
DOS in the insulating phase
is really a pseudogap, and is nonzero (albeit exponentially small)
in a region around the pseudogap.  Hence, the MIT on a hypercubic 
lattice always occurs when the self energy develops a pole (that lies 
infinitesimally below the real axis).  It turns out that
the same scenario occurs on the Bethe lattice at half filling, even though the 
DOS has a finite bandwidth---the MIT occurs when the self energy develops a pole
at the chemical potential (although here there is now a 
well-defined gap, where the DOS vanishes over a finite range of frequencies).

The question we wish to address is what happens when the MIT occurs for 
a system that does not possess
particle-hole symmetry~\cite{hirsch}.   Following the argument given 
above, the pole formation must be the underlying cause of the MIT on the 
hypercubic lattice, but the situation on the Bethe lattice is unclear. 
The interest in examining particle-hole asymmetric cases lies in the fact
that most 
real materials do not have particle-hole symmetry, so understanding
consequences of breaking particle-hole symmetry is important for understanding
experimental systems.  There are two ways to break particle-hole symmetry:
(i) one can modify the lattice, so it is not bipartite, then there is no 
particle-hole symmetry at half filling, where the MIT occurs, or (ii) one
can modify the model so that a MIT occurs at different fillings, away from
the particle-hole symmetric limit.  We choose to examine the latter here.

We consider the MIT in the spinless Falicov-Kimball 
(or simplified Hubbard) model \cite{falicov_kimball}, which is believed to 
describe correlated-electron behavior, and the MIT in particular, 
in materials that can be fit into a binary alloy picture. 
The canonical system that fits this picture is Ta$_x$N \cite{Kaul,Yu}
which has its MIT occur at a particle-hole asymmetric value of $x=0.6$.
The Falicov-Kimball model has been previously applied to the MIT 
\cite{vanDongen,Kalinowski}, its advantages over the Hubbard model
are that, it has a MIT for a wide range of  fillings and the numerics are
under much better control. 
The Falicov-Kimball Hamiltonian has the following form
\begin{equation}
H=-t\sum_{\langle i,j\rangle}c^\dagger_{i}
c_{j}+U\sum_{i}w_ic^{\dagger}_{i}c_{i},
\label{eq: ham}
\end{equation}
where $c^{\dagger}_{i}$ ($c_{i}$) denotes the creation (annihilation)
operator for a spinless electron on site $i$, and the summation is restricted to
nearest neighbors. The classical variable $w_i$ equals one or zero, corresponding to the 
presence of an A or B ion at the given lattice site, and $U$ is the diagonal
site-energy difference
between the two ionic configurations. 
The hopping integral $t$ is appropriately scaled in order 
to be finite in the limit of large dimensions ($d$) or coordination 
number ($Z$)~\cite{metzner_vollhardt},
and to result in the same effective bandwidths $W=\sqrt{\int{\epsilon^2 \rho({\epsilon}) d\epsilon}}$ (with $\rho(\epsilon)$ the noninteracting DOS)
for both lattices~\cite{NRG_Bulla}. 
Therefore, $t=t^*/\sqrt{Z}$ on the Bethe lattice and $t=t^*/\sqrt{2d}$ 
on the hypercubic lattice. 
The conduction electrons interact with the localized particles 
(which have an average filling of $w_1=\langle w_i\rangle$) with an interaction
strength $U$;
this forms the canonical binary alloy picture.
Such a picture is particularly useful because the model exhibits a MIT 
when the particle-hole symmetry is broken ($w_1 \ne 0.5$), as long as the
total particle density equals one.
Hence, we constrain our calculations to fix the total number
of particles, i.e. the number of conduction electrons satisfies $\rho_e=1-w_1$. 
If we choose to measure all energies in the units of $t^*=1$, then 
the noninteracting DOS on the Bethe and hypercubic lattices are
a semicircle and a Gaussian, respectively, $\rho_{Bethe}(\epsilon)=\sqrt{4-\epsilon^2}/2\pi$ and
$\rho_{HC}(\epsilon)=\exp(-\epsilon^2/2)/\sqrt{2\pi}$. 

In the limit of infinite dimensions the Falicov-Kimball model can be solved exactly using 
DMFT \cite{Brandt,freericks_review}. 
Because the self-energy $\Sigma(\omega)$ has no
momentum dependence one can employ an iterative scheme using the following relationships 
between the self-energy $\Sigma(\omega)$, the retarded 
Green's function $G(\omega)$, and the effective medium $G_0(\omega)$
\begin{equation}
G(\omega)=\int{d\epsilon \rho(\epsilon)\frac{1}{\omega+\mu-\Sigma(\omega)-\epsilon+i \delta }}
\label{G}
\end{equation}
\begin{equation}
G_0(\omega)=[G(\omega)^{-1}+\Sigma(\omega)^{-1}]^{-1}
\label{G_0}
\end{equation}
\begin{equation}
G(\omega)=(1-w_1)G_0(\omega)+w_1\frac{1}{G_0(\omega)^{-1}-U}
\label{G2}
\end{equation}
closed in the iterative loop until the self-energy is converged to a desired accuracy~\cite{jarrell}. 
The algorithm yields reliable results for relative accuracies up to $10^{-13}$. 
The calculations are particularly easy in the case of the Bethe lattice since the integral in 
Eq.~(\ref{G}) can be determined analytically (for the hypercubic lattice it is a complex error function).
Once the algorithm is converged, the interacting DOS is defined to be
$\rho_{int}(\omega)=-\textrm {Im}[G(\omega)]/\pi$.
The essential difference
in the physics on these two lattices arises from the behavior of the interacting DOS. 
While the Bethe lattice exhibits a well defined gap where the interacting DOS is 
exactly zero  for a finite range of frequencies when $U>U_{cg}$ (the critical
value of the interaction strength for opening a gap in the interacting DOS),
on the hypercubic lattice the DOS splits into two subbands divided by a ``pseudo-gap'', where the interacting 
DOS is exponentially small and exactly zero only at one point. 
The interacting DOS 
away from half filling has appeared elsewhere \cite{vanDongen,our_hyper}. 
We should note that in the Falicov-Kimball model there is no quasiparticle peak 
at the chemical potential, which normally develops for $U$ less than $U_{cg}$ 
in the Hubbard model in infinite dimensions. This is because the Falicov-Kimball
model lacks quasiparticles in the strict definition of the term, since the 
lifetime of Fermionic excitations never becomes infinite as $T\rightarrow 0$. 

\begin{figure}[htbf]
\epsfxsize=3.0in
\centerline{\epsffile{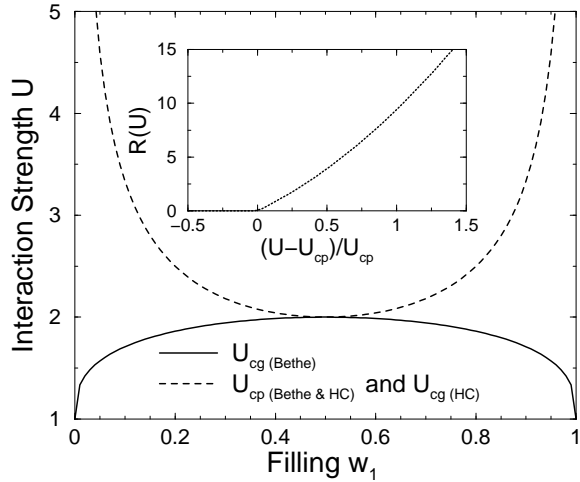}}
\caption{Phase diagram for the MIT in the Falicov-Kimball model on the Bethe
and hypercubic lattices. The solid line denotes $U_{cg}$ for the gap formation
on the Bethe lattice, the dashed line denotes $U_{cp}$ for the pole formation
on the Bethe lattice and $U_{cg}=U_{cp}$ on the hypercubic lattice. The inset 
shows the universal curve of the residue of the pole as a function of the 
interaction strength $R(U)$ for both lattices and all fillings. 
\label{pole_gap}}
\end{figure}

The metal-insulator transition at half-filling is closely related to the development
of a pole in the self-energy, where the DOS is suppressed to zero ($U_{cg}$)
at the same interaction strength  
for which the pole forms ($U_{cp}$); i.e., $U_{cg}=U_{cp}$. Therefore one could suggest to use the residue of the 
pole as an order parameter for the MIT. In fact, a plot of the residue versus
$(U-U_{cp})/U_{cp}$ is universal, for all fillings on both the hypercubic
and Bethe lattices, indicating that a scaling theory holds for the residue
of the pole (see the inset to Fig.~\ref{pole_gap}).
Away from half-filling, these two processes (pole formation and the MIT)
are decoupled on the Bethe lattice, with the pole formation occurring after
the MIT (in particular, the real part of the Green's function does not cross
the horizontal axis [within the band gap] until the pole forms).  

We begin with the cubic equation 
which is satisfied by the Green's function on the Bethe lattice \cite{vanDongen}
\begin{equation}
G^3-2xG^2+(1+x^2-\frac{U^2}{4})G-(x+\alpha)=0
\label{cubic}
\end{equation}
where $x\equiv\omega+\mu-U/2$ and $\alpha\equiv U(w_1-\frac{1}{2})$.
The pole in the self-energy 
develops when $G=0$, requiring $x+\alpha=0$, or $\omega_{pole}=U(1-w_1)-\mu$. 
Next, one notes that $G=0$ is a physical root~\cite{physical}
when $1+x^2-U^2/4\le 0$, which 
leads to the simple formula for $U_{cp}$, the critical value of $U$ at which the 
self-energy develops a pole
\begin{equation}
U_{cp}=\frac{1}{\sqrt{w_1(1-w_1)}}.
\label{U_cp}
\end{equation}
Away from half filling, the critical values of $U$ for the 
MIT and the pole formation are not necessarily the same on the Bethe lattice, 
and the former is found from 
the following. Let us use standard notation for the cubic equation's coefficients (see \cite{stegun})
$q\equiv(3-x-3U^2/4)/9$ and $r\equiv x(1+U^2/2)/6+\alpha/2-x^3/27$. 
The condition for the location of the
band edges ($q^3+r^2=0$)  is a fourth order equation in 
$\omega$, and must have exactly three distinct roots at $U_{cg}$ the critical 
interaction strength for the MIT
(four distinct roots for $U>U_{cg}$). As shown 
in Ref. \cite{vanDongen} this condition leads to an equation for
the critical interaction:
\begin{equation}
1-4\frac{\alpha^2}{U^2}=\frac{4(U^2-1)}{27U^2}.
\label{gap_condition}
\end{equation}
Solving Eq. (\ref{gap_condition}) 
for $U$ yields the critical interaction strength for the MIT (gap opening
on a Bethe lattice) $U_{cg}$:
%
%
\begin{equation}
U_{cg} = \sqrt{1+3w_1^{1/3}(1-w_1)^{1/3}[(1-w_1)^{1/3}+w_1^{1/3}]}.
\label{U_cg}
\end{equation}
Both expressions (\ref{U_cp}) and (\ref{U_cg}) yield 2 at half filling
($w_1=0.5$). The behavior of $U_{cp}$ and $U_{cg}$ for arbitrary fillings
is shown in Fig.~\ref{pole_gap}.
As the system moves away from half-filling the particle-hole asymmetry 
allows for the formation of a third ``phase'', 
in which the interacting DOS has a gap but there is no pole in the self energy
(region between the solid and dashed lines in Fig.~\ref{pole_gap} on the Bethe
lattice).  This dramatic difference in the MIT as we move away from half
filling (where the transition occurs at smaller $U$ values than that at half 
filling on the Bethe lattice, but larger $U$ values on the hypercubic lattice)
is likely tied to the fact that there is an infinite number of neighbors
and loops on a hypercubic lattice, which allows one to avoid sites blocked
by the presence of a static particle, but on the Bethe lattice, whole subtrees
become disconnected because there are no loops, so the blockage is more 
effective.  
On the hypercubic lattice, the
pole formation and MIT always occur at the same value of $U$ ($U_{cg}=U_{cp}$), 
and it is possible to obtain an exact expression for the pole formation, 
which turns out to coincide with Eq. (\ref{U_cp}).
Numerical calculations show that the interacting DOS for the hypercubic lattice 
has an exponentially small region over a similar range in frequency as the gap 
region on the Bethe lattice (for the same value of $U$). 

\begin{figure}[htbf]
\epsfxsize=3.0in
\centerline{\epsffile{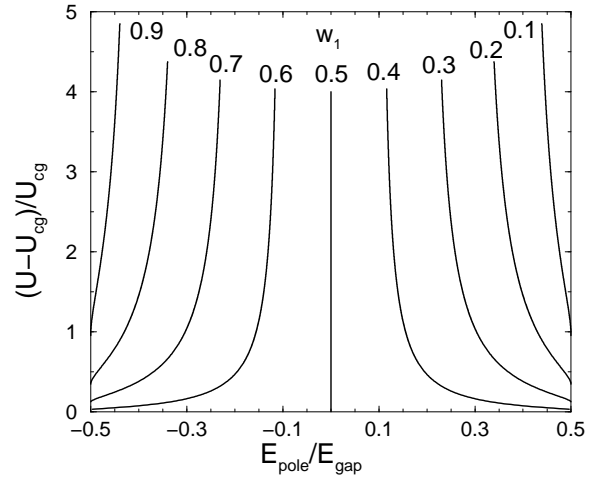}}
\caption{The relative interaction strength for the opening of a gap versus 
the relative location of the pole ($E_{pole}$) in the gap (of width $E_{gap}$)
on the Bethe lattice; the origin on the horizontal axis lies at the center
of the gap. The lines from left 
to right correspond to different fillings $w_1$ ranging from 0.9 to 0.1, 
in steps of 0.1. 
\label{U_c_gap}}
\end{figure}

Fig.~\ref{U_c_gap} shows the relative interaction strength
as a function of the relative location of the pole 
within the gap, on the Bethe lattice. The pole 
is located in the middle of the gap at half filling for any value of $U>U_{cg}$. However, as the 
particle-hole symmetry is broken, the pole first appears at the lower or upper
band edge (for $U=U_{cp}$), depending
on whether $w_1$ is larger or smaller than 0.5, and as $U$ increases the pole 
drifts closer to the center of the bandgap. Note that there is no smooth transition 
between the half-filled case and the particle-hole asymmetric case. 

\begin{figure}[htbf]
\epsfxsize=3.0in
\centerline{\epsffile{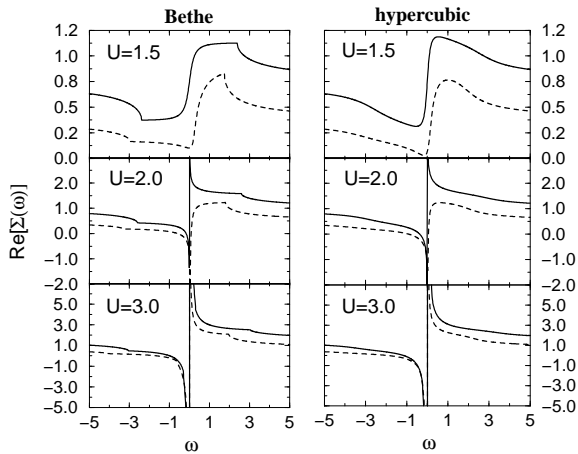}}
\caption{Evolution of the real part of self-energy for different values of $U$, 
for both the Bethe (left panel) and hypercubic (right panel) lattices, at half 
filling $w_1=0.5$ (solid line) and at $w_1=0.25$ (dashed line). The origin of 
the frequency axis is set to $U(1-w_1)-\mu$.
\label{Evolution}}
\end{figure}

Numerical calculations show the evolution of the real part of self-energy on 
the Bethe lattice compared to that on 
the hypercubic lattice in Fig.~\ref{Evolution}, when approaching the MIT from the metallic side. 
Aside from obvious similarities between the two 
lattices, this graph demonstrates the existence of the third phase in the middle panel, where $U=2$. 
The half-filled curves ($w_1=0.5$) always show poles in the insulator, and therefore, the  DMFT scenario discussed in 
Refs. \onlinecite{ipt,NRG_Bulla,AntiKotliar} holds.
However,  
the curves corresponding to $w_1=0.25$ exhibit large (negative) but finite values of 
$\textrm{Re}[\Sigma(\omega)]$ (the pole has not yet developed) but the system is
an insulator with a well developed gap at this value of $U$.
Hence, we conclude that the development of the pole and the MIT 
are decoupled away from half filling and even if it might be tempting
to use the residue of the pole as an ``order parameter'' for the MIT, it
fails to describe the situation off of half filling on lattices with a
finite bandwidth.  So what significance can be made of the pole formation?
In order to investigate this, we 
calculate the dc conductivity in the 
relaxation time formalism (see \cite{Freericks_Mahan_theorem} for details). 
These calculations (not shown here) indicate that  the MIT is always 
a continuous transition at $T=0$, with the conductivity being suppressed
continuously to zero as $U\rightarrow U_{cg}$.
Also, at finite temperature we see no evidence for the influence of the
pole on the transport.  The conductivity curves are smooth
functions of $T$ with no unusual features occurring when $U>U_{cp}$.

We need to note that in the correlated insulator, in the vicinity of $\omega=U(1-w_1)-\mu$, one can obtain 
analytic expressions for the Green's function $G(\omega)$ and the self-energy 
$\Sigma(\omega)$ on the hypercubic lattice \cite{our_hyper}. 
As expected, one obtains imaginary parts of $G(\omega)$ and $\Sigma(\omega)$ 
exponentially approaching 
zero as $\omega \rightarrow U(1-w_1)-\mu$, while they are exactly zero on the Bethe lattice.
These, at first sight, small differences, in fact turn out to cause dramatic changes 
in the low temperature transport properties \cite{our_hyper,our_bethe} when going from the Bethe lattice (where the relaxation time vanishes in the gap)
to the hypercubic lattice (where the relaxation time has a power-law dependence
not an exponentially decaying dependence around the pseudogap), and therefore 
must be dealt with carefully. The
most significant departure is in thermal transport properties, where the
thermopower diverges on the Bethe lattice, but vanishes on the hypercubic
lattice in the insulating phase as $T\rightarrow 0$ (when particle-hole
symmetry is broken).

In conclusion, we have analyzed the effect of particle-hole asymmetry on
the Mott transition  
in the infinite dimensional Falicov-Kimball model (on both the Bethe and the 
hypercubic lattices).
Hitherto, it was believed that the scenario for the MIT on both lattices was the same, as indeed is the case at half-filling.
We find that this is not true when the particle-hole symmetry is removed, as is often the case in real materials.
We show that in the absence of particle-hole symmetry the pole formation and the MIT
are two unrelated processes on the Bethe lattice. So even though the residue
of the pole satisfies many of the properties expected of an order parameter,
it cannot be employed to describe the MIT in all cases (although one could
use it on the hypercubic lattice).  Furthermore, there seems to be little 
difference between the properties of a correlated insulator with or without
a pole in the self energy. We conjecture that all of the conclusions about the
character of the MIT on the Bethe lattice will hold for other systems
with a finite bandwidth, and hence much of these results will play a role
in realistic models of the MIT for real materials in finite dimensions.


We would like to acknowledge useful discussions with V. Zlati\'c, V. Turkowski, and A. G. Petukhov. 
This work was supported by the National Science Foundation, 
grant DMR-0210717, and by the Office of Naval Research, grant 
number N00014-99-1-0328. 



\begin{references}
\bibitem{physical} 
The physical root satisfies $\textrm{Im} [G] < 0$; when all roots are real, 
one needs to use continuity to determine the physical branch. When the self
energy has a pole corresponding to $G=0$, the root is physical when 
$\textrm{Im} [G] <0$ for $\omega$ lying infinitesimally above the real axis.
\end{references}
\end{document}